\newif\ifarxiv
\arxivtrue            

\ifarxiv
  \documentclass[sigconf,nonacm]{acmart}
\else
  \documentclass[sigconf,anonymous]{acmart} 
\fi

\renewcommand\footnotetextcopyrightpermission[1]{}
\setcopyright{none}
\ifarxiv
\else
\fi

\usepackage[noend]{algpseudocode}
\algrenewcommand\algorithmiccomment[1]{\hfill\texttt{// #1}}
\usepackage{booktabs}
\usepackage{tabularx}
\usepackage{array}
\usepackage{float}
\usepackage{tikz}
\usetikzlibrary{shapes.geometric}
\newcommand{\circled}[1]{\tikz[baseline=(char.base)]{
    \node[shape=circle,fill=black,text=white,inner sep=1pt] (char) {#1};}}

\newcommand{\triyellow}{\tikz[baseline=-0.3ex]{
    \node[regular polygon, regular polygon sides=3, fill=yellow,
          draw=black, inner sep=1.5pt, minimum size=0.8em] {};}}

\newcommand{\trigreen}{\tikz[baseline=-0.3ex]{
    \node[regular polygon, regular polygon sides=3, fill=green,
          draw=black, inner sep=1.5pt, minimum size=0.8em] {};}}

\newcommand{\trired}{\tikz[baseline=-0.3ex]{
    \node[regular polygon, regular polygon sides=3, fill=red,
          draw=black, inner sep=1.5pt, minimum size=0.8em] {};}}

\author{Yihe Zhang$^*$}
\affiliation{%
  \institution{University of Illinois Chicago}
  \city{Chicago}
  \state{Illinois}
  \country{USA}
}
\email{yihe6@uic.edu}

\author{Yash Kurkure$^*$}
\affiliation{%
  \institution{University of Illinois Chicago}
  \city{Chicago}
  \state{Illinois}
  \country{USA}
}
\email{ykurku2@uic.edu}

\author{Yiheng Tao}
\affiliation{%
  \institution{University of Illinois Chicago}
  \city{Chicago}
  \state{Illinois}
  \country{USA}
}
\email{ytao28@uic.edu}

\author{Michael E. Papka}
\affiliation{%
  \institution{University of Illinois Chicago}
  \city{Chicago}
  \state{Illinois}
  \country{USA}
}
\affiliation{%
  \institution{Argonne National Laboratory}
  \city{Lemont}
  \state{Illinois}
  \country{USA}
}
\email{papka@anl.gov}

\author{Zhiling Lan}
\affiliation{%
  \institution{University of Illinois Chicago}
  \city{Chicago}
  \state{Illinois}
  \country{USA}
}
\affiliation{%
  \institution{Argonne National Laboratory}
  \city{Lemont}
  \state{Illinois}
  \country{USA}
}
\email{zlan@uic.edu}

\begin{document}

\title{A Real-Time Digital Twin for Adaptive Scheduling}

\begin{abstract}
High-performance computing (HPC) workloads are becoming increasingly diverse, exhibiting wide variability in job characteristics, yet cluster scheduling has long relied on static, heuristic-based policies. 
In this work we present SchedTwin, a real-time digital twin designed to adaptively guide scheduling decisions using predictive simulation.
SchedTwin periodically ingests runtime events from the physical scheduler, performs rapid what-if evaluations of multiple policies using a high-fidelity discrete-event simulator, and dynamically selects the one satisfying the administrator configured optimization goal. 
We implement SchedTwin as an open-source software and integrate it with the production PBS scheduler. 
Preliminary results show that SchedTwin consistently outperforms widely used static scheduling policies, while maintaining low overhead (a few seconds per scheduling cycle). These results demonstrate that real-time digital twins offer a practical and effective path toward adaptive HPC scheduling. 
\end{abstract}


\maketitle

\def\thefootnote{*}\footnotetext{equal contribution}\def\thefootnote{\arabic{footnote}}

\section{Introduction}

\begin{figure}
    \centering
    \includegraphics[width=0.8\linewidth]{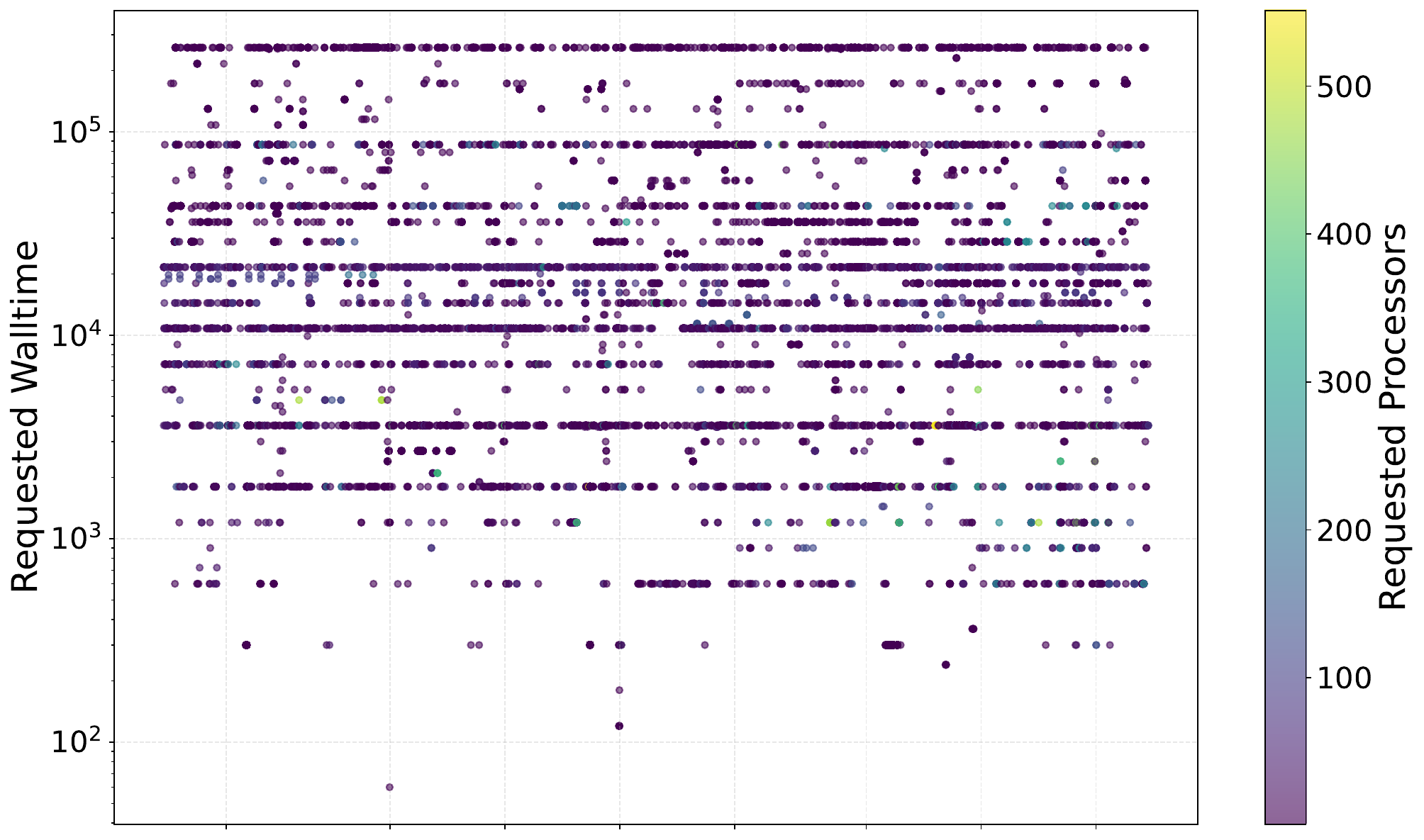}
    \caption{Job distribution on the Polaris system at ALCF (January 18 - March 18, 2024).  }
    \label{fig:polaris_log}
\end{figure}

In high-Performance Computing (HPC),  workloads are becoming increasingly diverse, with variable job characteristics.
For example, Figure~\ref{fig:polaris_log} shows the job distribution on the Polaris system at Argonne Leadership Computing Facility~\cite{ALCFDocs}, revealing substantial variations in both job sizes and runtimes.
Resource management and scheduling (RMS) systems, also known as \emph{cluster schedulers} or \emph{batch schedulers}, are responsible for selecting users jobs from job queues for execution, typically in bare-metal mode with exclusive node access.
Thus, cluster schedulers are crucial to determining both the time-to-solution for users and the utilization of million-dollar systems. 

Over the past decades, cluster scheduling has relied primarily on \emph{heuristic-based methods} \cite{FCFS_BF, jsspp2017, slurmOverviewSchedMD, LLNL_Flux_Framework}. For example, First-Come-First-Served (FCFS) and utility-based scheduling policies, often augmented with backfilling, are commonly deployed to manage HPC resources \cite{nersc, ALCFQueueScheduling, olcfFrontierSchedulingPolicy, tacc}. 
FCFS scheduling executes jobs according to their arrival times, while utility-based scheduling chooses the job that maximizes a specific utility value, often defined by job attributes.
Due to their reliance on fixed scheduling strategies, these conventional policies are inherently \emph{static} and cannot adapt to evolving workload characteristics or system state. Furthermore, their \emph{one-step decision-making process} provide no guarantees of near-optimal performance.
Consequently, the inability of conventional scheduling approaches to deliver near-optimal performance for dynamic workloads has emerged as a critical limitation for future cluster scheduling.

Recently, the community is actively exploring the concept of a Digital Twin (DT) for modeling complex systems. A DT is defined as a virtual, high-fidelity counterpart of a physical system. 
For example, the ExaDigit project has developed digital twins to represent various supercomputer architectures\cite{exadigit_2}.
While these efforts have made notable contributions, existing work predominantly uses DTs driven by historical log data to analyze policy behavior retrospectively, such as for performance analysis, what-if scenario exploration, or design-time evaluation of future system prototypes, rather than to influence live scheduling operations. 
\emph{This work aims to investigate a digital twin designed for guiding adaptive scheduling in real time.}

We introduce \emph{SchedTwin}, a real-time digital twin (DT) framework that periodically observes the live system state, rapidly evaluates the long-term outcomes of multiple scheduling policies, and dynamically selects optimal policy in response to evolving system conditions.
Here, \emph{real time} refers to the ability to make dynamic scheduling decisions within seconds.
In contrast to conventional heuristics, SchedTwin offers \emph{two distinctive advantages}. First, SchedTwin is \emph{intelligent} because it leverages high-fidelity simulation to evaluate the long-term performance of different policies before making a decision.  This capability enables SchedTwin to optimize future scheduling outcomes rather than focusing solely on the immediate next step.
Second, instead of relying on a fixed strategy, SchedTwin is \emph{adaptive}, dynamically selecting the most appropriate scheduling policy in response to changing system conditions, such as evolving workload characteristics and resource availability.

Overall,  this work makes three key contributions: (i) developing SchedTwin, a simulation-in-the-loop digital twin that closes the feedback loop with a production scheduler for adaptive scheduling in HPC; (ii) implementing SchedTwin as open-source software; and (iii) demonstrating SchedTwin’s promising performance as a real-time digital twin.

\section{Related Work}

\subsection{Cluster Scheduling in HPC}
Cluster scheduling, also known as resource management and scheduling (RMS), is an indispensable system software that mediates between users and hardware resources. It assigns jobs to compute resources (e.g., compute nodes) according to a site policy and resource availability \cite{jsspp2017}. Well-known schedulers include Slurm, PBS, and Flux~\cite{slurmOverviewSchedMD, altairPBS2022, LLNL_Flux_Framework}.

In HPC systems, users submit jobs to queues; jobs are dispatched as resources become available and are executed in bare-metal mode with exclusive node access. As a result, user jobs receive the full allocation of CPU, memory, and other hardware resources on their assigned nodes, while the scheduler must arbitrate resource usage across multiple queued jobs to maintain both high resource utilization and user-level fairness.

HPC systems typically employ static, heuristic-based scheduling policies, often combined with queue configurations and backfilling\cite{alcfAuroraRunningJobs,olcfFrontierSchedulingPolicy,nerscQueuesChargesPolicy,tacc}. First-Come-First-Serve
(FCFS) with EASY backfilling, a widely used method, sorts the jobs in the wait queue by arrival time and schedules them from the head of the queue\cite{FCFS_BF}. When insufficient resources are available for the first job, the scheduler reserves resources for it and uses backfilling to improve utilization, allowing later jobs to run as long as they do not delay existing reservations. 
Another widely adopted approach, broadly referred to as utility-based policy, orders jobs according to a utility function that depends on job attributes (e.g., size, runtime estimates), time in queue, and resource usage, especially in leadership computing facilities\cite{jsspp2017}.

In contrast to static heuristics that lack adaptability and rely on local decision-making, SchedTwin uses rapid predictive simulation to dynamically select the most suitable scheduling policy based on current job and resource states, offering an adaptive and intelligent approach.

\subsection{Digital Twins}
The concept of Digital Twins (DTs) for computer systems has evolved significantly, from early standalone simulation tools to  data-driven virtual replicas of full machines and even entire data centers. Recent state-of-the-art work, such as the ExaDigiT framework, demonstrates DTs that integrate resource allocation, power conversion, cooling, and advanced visualization to study efficiency, reliability, and operational strategies \cite{matthias_dt_sc24,matthias_dt_sc25w} entirely in silico. 
While these efforts provide significant value, their primary focus is on offline planning, e.g., exploring optimizing capacity management or validating design-time evaluations for future system prototypes, rather than providing real-time feedback for dynamic operational control.

The community typically employs discrete-event simulation (DES) for modeling and analyzing scheduling processes\cite{more_for_less, GridSim, AccaSim, batsim, symbiotic_sim}.
DES models a scheduling process as a sequence of instantaneous events, such as job arrivals, job starts, and job completions, that update system state. A scheduling instance is triggered when a job arrives or a running job completes; the scheduler then iterates over queued jobs and allocates resources according to the configured policy. 
Time advances by ``jumping''  from one event to the next rather than evolving continuously, making DES a natural fit for modeling job queuing and resource allocation. 

Distinguishing from offline-oriented use of DTs, \emph{SchedTwin} emphasizes a real-time digital twin that operates in the loop with a production scheduler to guide scheduling decisions. It leverages DES for predictive simulation, in which real-time predictive analysis is used to steer decisions in the physical system~\cite{Symbiotic_simulation_2002,symbiotic_simulation,Portfolio}.

\section{SchedTwin Design}
\label{sec:overview}

\emph{SchedTwin} is a real-time digital twin that operates alongside a production scheduler to guide adaptive scheduling. 
At its core, SchedTwin includes a predictive simulator that dynamically selects an appropriate policy at each scheduling interval based on an operator-specified objective and the current scheduling state, and then returns this decision to the physical scheduler as actionable feedback. The design of SchedTwin is general, as it can easily incorporate a pool of candidate policies, such as widely deployed static scheduling policies, provided that they exhibit complementary strengths.

\begin{figure*}[htbp]
  \centering
  \includegraphics[width=0.8\linewidth]{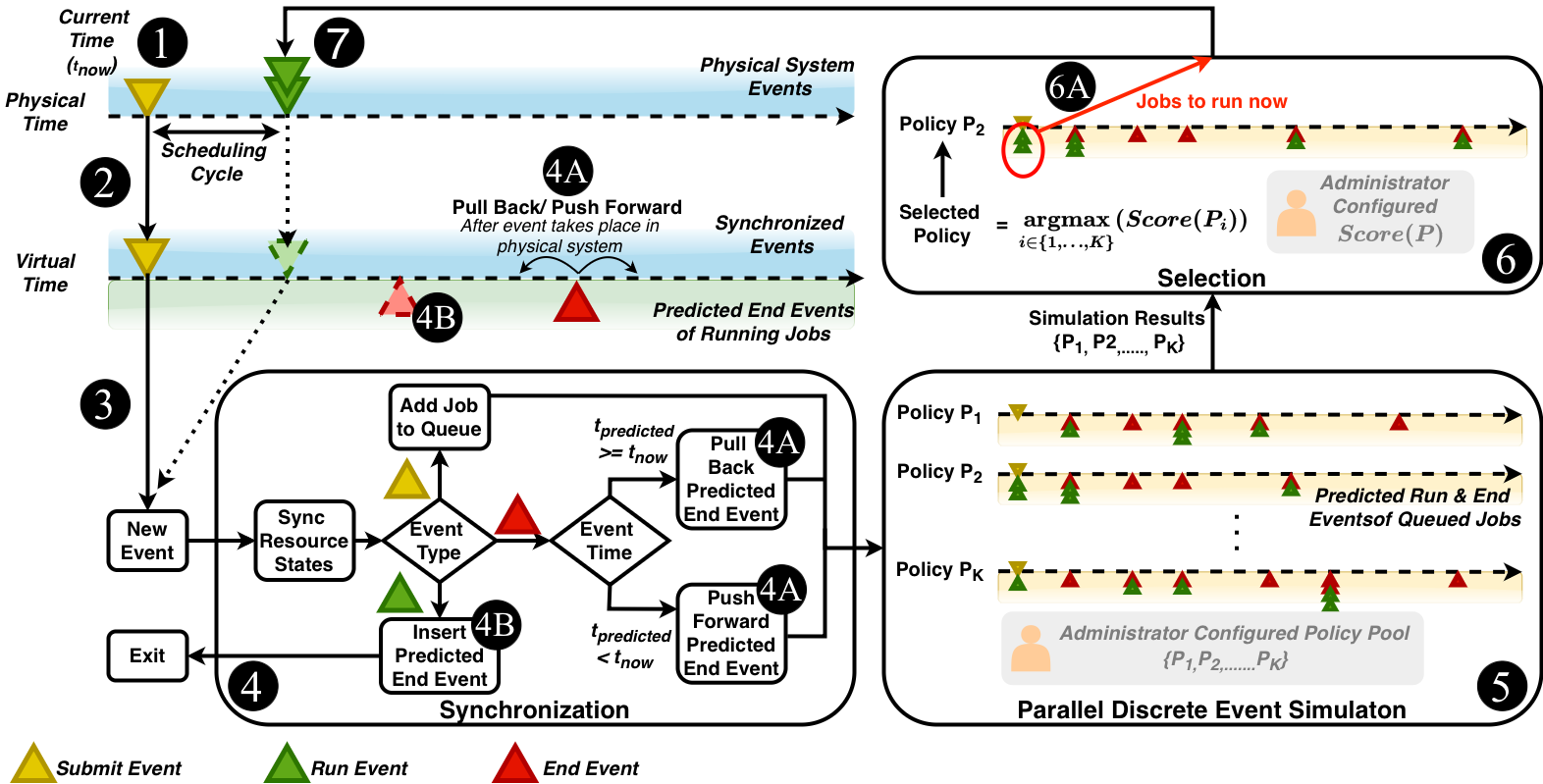}
  \caption{Workflow of SchedTwin. The scheduler triggers SchedTwin on new job and system events; SchedTwin synchronizes its internal view with the scheduler, issues predictive simulations based on the updated events, and returns informed scheduling decisions to the scheduler.}
  \label{fig:timeline}
\end{figure*}

Figure \ref{fig:timeline} presents the SchedTwin's workflow where the top right visualizes two timelines: \textbf{Physical Time} denotes the physical system's scheduler timeline and \textbf{Virtual Time} denotes SchedTwin's timeline. 

\subsection{Event Streaming}
\label{sec:event_stream}

SchedTwin is triggered by job- or resource-related events, such as a user submitting a job, the scheduler starting a job, a running job completing, or other job/resource changes.
Every scheduling cycle begins at \circled{1}, where an event takes place in the physical system's scheduler. This can either be a job submit event (\triyellow), job run event(\trigreen) or job end event(\trired) as shown in the figure. 

In SchedTwin, \emph{a lightweight in-memory Redis data store} is deployed to stream events from production schedulers to the DT: the scheduler acts as the producer and SchedTwin as the consumer of these streams. 
We develop custom scripts that are triggered by scheduler events, such as PBS-defined \texttt{queuejob}, \texttt{runjob}, and \texttt{jobobit} events. The scripts collect real-time metadata from the scheduler and publish it to a Redis stream at \circled{2}. Finally, SchedTwin subscribes to this event stream at \circled{3}, reads incoming events, and synchronizes its internal state accordingly.

\subsection{Synchronization}
\label{sec:synch}

Upon receiving streamed events from the scheduler, SchedTwin checks whether an event is a job submission or completion event, since scheduling instances are
triggered when new jobs enter the queue or when jobs finish and free resources. A scheduling cycle begins when a scheduling decision is required. SchedTwin uses predictive simulation to determine which jobs it may run next.

In Figure \ref{fig:timeline} the virtual timeline beyond the current time depicts the event queue of the simulator. The future events on the time line can only be job end events (\trired), because job submit events (\triyellow) cannot be predicted and job run events (\trigreen) can only be inserted into the event queue when a job actually runs on the physical system. The future job end events (\trired) are for the jobs currently running on the HPC system and their events are queued based on user estimated wall time. 

When a new event occurs, the control goes to the synchronization block \circled{4}. It first synchronizes the current system status including node availability information using command-line tools. Then, the type of the event is checked. If it is a job run event (\trigreen), a job end event (\trired) is inserted into the event queue \circled{4B} and SchedTwin exits immediately as run events don't imply new scheduling opportunities. If it is a job end event (\trired), SchedTwin must carefully synchronize the system state 
due to commonly inaccurate user-provided wall times which
can cause jobs to finish earlier than predicted.
In addition, a job may appear to finish later than predicted if it takes time for the scheduler 
to clean up resources after job completion. 
If the job ends earlier or later, the predicted end time is pulled back or pushed forward respectively as shown in \circled{4A}.
The latter is uncommon because users typically overestimate runtime and cleanup delays are usually small. Lastly, if it is a job submit event (\triyellow), the job is added to the queue of waiting jobs. Apart for the job run event, the other two events lead to the next stage of parallel discrete event simulation as a new job or freed resources indicate possible scheduling opportunities. 

\subsection{Parallel Discrete Event Simulation}
\label{sec:para-sim}

After synchronization, SchedTwin employs predictive simulation for its scheduling decision \circled{5}.  
We parallelize the predictive simulation such that all what-if explorations for each policy run in parallel. This ensures that the overhead never exceeds the runtime of the longest running simulation.

Our predictive simulator is built on CQSim\cite{CQSimRepo}. 
CQSim is a discrete event simulator for scheduling, and has been extensively used in the past ~\cite{yang_sc13_dynamic_pricing}. 
Specifically, the predictive simulator models the scheduling system as a sequence of job events (e.g., submission, start, completion) and resource events (e.g., allocation, deallocation, failures), where each event occurs at an instant in time and triggers a change of system state. 

In this stage, $k$ copies of the simulator object are created. This does not incur a large overhead as all objects share a common database and only carry event metadata required for simulation. Each copy of the simulator is then configured with a single policy $P_i \in {P_1,P_2,...P_k}$ which is the policy pool configured by the administrator. Each simulator is then run until no jobs are left in the queue. This predictive simulation facilitates the what-if exploration between the various policies in the policy pool. Each simulation runs in parallel independent of each other and return the final simulator object to the master process. These simulator objects are then passed on to the policy selection stage as simulation results.

\subsection{Policy Selection}
\label{sec:selection}

The policy selection state \circled{6} is where the best policy is chosen, enabling the adaptability of SchedTwin. At this stage, the simulation result for each policy $P_i \in {P_1,P_2,...P_k}$ is passed to a scoring function, $Score(P_i)$, which is configured by the system administrator. 
This function may incorporate various metrics such as wait time, job slowdown, or resource utilization at the end of each simulation. 
The function produces a score for each policy, and the policy with the highest score is selected. 
The selected policy's simulator instance is then queried for the job run events \circled{6A} 
that occur immediately after the current time; these events correspond to the jobs that SchedTwin selects to run next.

\subsection{Decision Feedback}
\label{sec:feedback}

Once jobs are selected to run by SchedTwin, The feedback can be delivered via the interfaces provided by production schedulers, allowing SchedTwin to inform scheduling decisions seamlessly without modifying the core scheduler logic. For example, PBS’s \texttt{qrun <jobid>} command can be used to run these jobs on the physical system \circled{7}. 

\subsection{Implementation}
\label{sec:imp}

We implement SchedTwin for use with the open-source version of the PBS scheduler \cite{openpbs}. Our implementation comprises approximately 1,000 lines of code extending the discrete-event scheduling simulator CQSim\cite{CQSimRepo}. 

\section{Evaluation}
\label{sec:results}

\subsection{Setup}
To evaluate real-time scheduling performance, we deploy SchedTwin on a 32-node cluster managed by PBS. The cluster is constructed using Docker containers running on an AMD x86 node with 48 cores, provisioned through CloudLab \cite{Cloudlab}. Of these containers, 32 represent compute nodes hosting PBS worker daemons, while one additional container hosts PBS administrative daemons, including the scheduler. We create a synthetic trace to test SchedTwin's adaptability. Specifically, the workload comprises 150 jobs in four phases: (1) a warm-up of 25 small jobs, each requesting 2–4 nodes with walltimes of 60–180 seconds; (2) a burst of 35 large, long jobs, each requesting 16–20 nodes with walltimes of 500–700 seconds; (3) a steady phase of 40 jobs, each requesting 6–8 nodes with walltimes of 200–300 seconds; and (4) a short-job tail of 50 jobs, each requesting 2–4 nodes with walltimes of 60–180 seconds. Job arrival rate is 5 seconds per job. 

We evaluate scheduling using both user-level and system-level metrics:\emph{job wait time}, \emph{job slowdown}, and \emph{system utilization}. To visualize overall scheduling performance, we use a Kiviat (radar) chart, reporting average and maximum values for wait time and slowdown alongside system utilization.  The area enclosed by a policy on the chart provides \emph{a holistic view} of its effectiveness, with a larger area corresponding to better overall performance.

We compare SchedTwin with three baselines:
(i) \textit{First-Come First-Serve with backfilling} (\textbf{FCFS})\cite{FCFS_BF},
(ii) \textit{Utility-based with backfilling}(\textbf{WFP}) (a utility-based policy used at ALCF \cite{jsspp2017}, 
and \textit{Short-Job-First}(\textbf{SJF}). In SchedTwin, the scheduling score, $Score(p)$, is defined as a weighted sum of key performance metrics:
\[
0.25\mathrm{maxWT}(p) + 0.25\mathrm{maxSD}(p) \\
  + 0.25\mathrm{avgWT}(p) + 0.25\mathrm{avgSD}(p).
\]
Here, $\mathrm{maxWT}(p)$ and $\mathrm{avgWT}(p)$ represent the maximum and average wait time respectively, while $\mathrm{maxSD}(p)$ and $\mathrm{avgSD}(p)$ represent the maximum and average slowdown achieved by policy $p$ across all jobs waiting in the queue. 

\subsection{Results}
\begin{figure}[htbp]
  \centering
    \includegraphics[
     width=0.8\linewidth,
        trim=0pt 8pt 0pt 0pt,
        clip
    ]{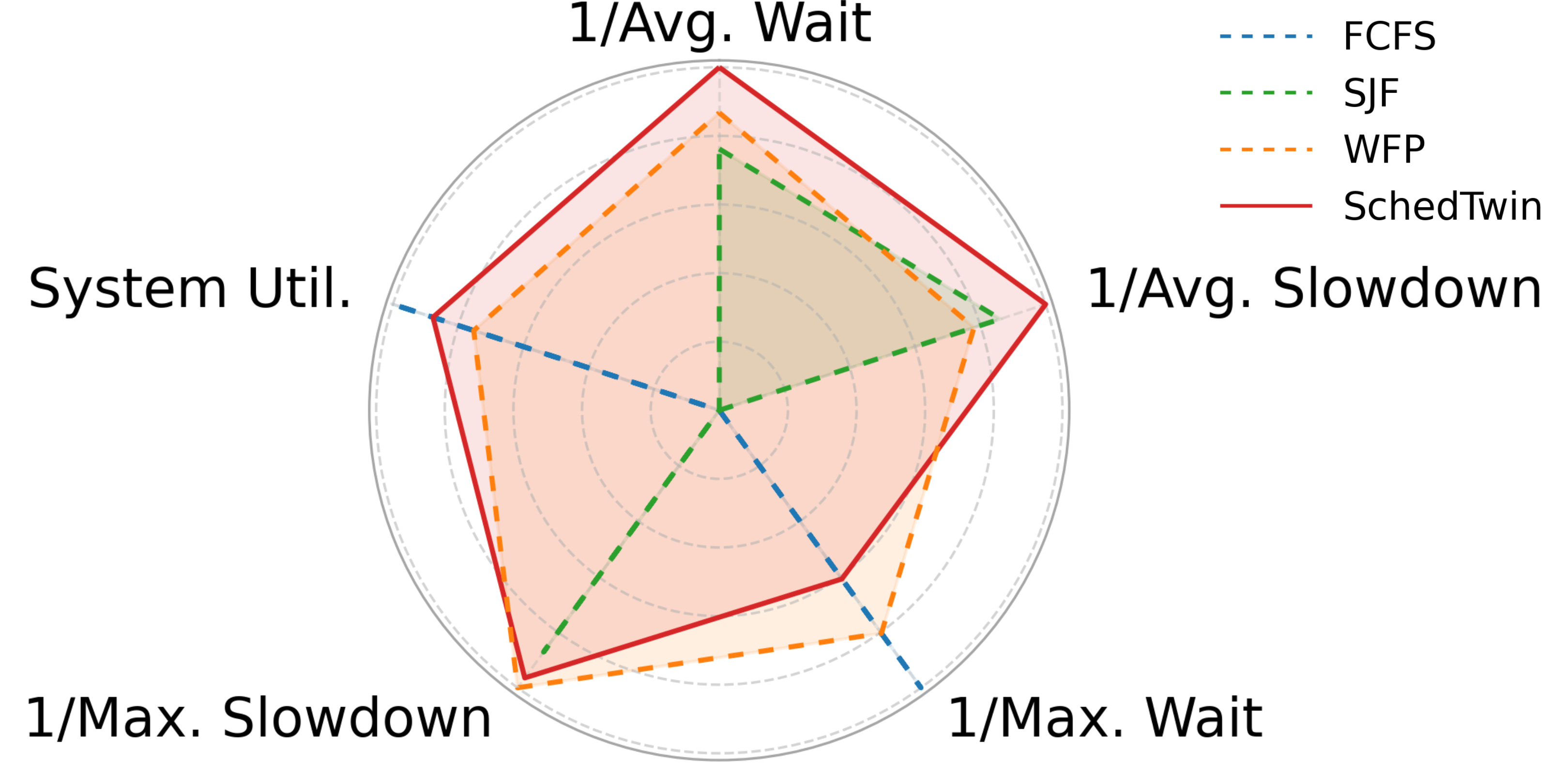}
    \caption{Scheduling performance comparison (synthetic workload). 
    A larger radar area indicates better overall performance. 
    Measured areas are FCFS: 0.00, SJF: 0.31, WFP: 1.67, and SchedTwin: 1.86.
    }
  \label{fig:synthetic_radar}
\end{figure}

Figure~\ref{fig:synthetic_radar} presents the radar chart comparing different scheduling methods on the synthetic workload. In this Kiviat plot, a larger covered area corresponds to better aggregate performance across all normalized metrics. Based on the reported statistics, SchedTwin achieves the best overall performance among the four methods, with an 11.4\% improvement over the second-best policy, WFP.

Because SchedTwin adaptively selects a policy rather than following a single static rule, it is useful to examine which policy actually initiates each job’s start. Table~\ref{tab:jobs-started-synthetic} reports the number of jobs started under each policy, revealing how SchedTwin leverages different policies to meet its scheduling objective. In some scheduling cycles, two or all three policies can attain the same objective score; in such ties, SchedTwin breaks the tie using the priority order WFP(utility-based policy)~$\rightarrow$~FCFS~$\rightarrow$~SJF, which also roughly reflects the relative popularity of these static policies in practice. Since tie cases are assigned to the selected policy via this rule, the table reports the resulting per-policy breakdown after tie resolution.

As discussed above, the synthetic trace is intentionally designed so that large, long jobs block subsequent short, small jobs, a situation where SJF is expected to be particularly effective. Consistent with this design, Table~\ref{tab:jobs-started-synthetic} shows that SJF is the static policy that most frequently attains the best objective value among the three candidates. However, to achieve the best overall balanced performance, SchedTwin still adaptively selects other policies when appropriate, since a purely SJF-driven strategy can sacrifice tail job latency, which is misaligned with the composite optimization objective. Calling back to the radar plot in Figure~\ref{fig:synthetic_radar}, we can see this trade-off visually: SJF achieves performance comparable to SchedTwin in terms of average waiting time and slowdown, but it is clearly the worst in system utilization and maximum waiting time, indicating that it severely penalizes long jobs. This again demonstrates that SchedTwin adaptively selects scheduling policies based on the current job mix in the queue and the specified composite objective.

\begin{table}[t]
  \centering
  \caption{Distribution of scheduling policies selected by SchedTwin on the synthetic workload. The values indicate the percentage of jobs started under each policy.}
  \label{tab:jobs-started-synthetic}
  \begin{tabularx}{\columnwidth}{@{}l*{3}{>{\centering\arraybackslash}X}@{}}
    \toprule
    & WFP & FCFS & SJF \\
    \midrule
    SchedTwin & 35.19\% & 15.66\% & 49.15\% \\
    \bottomrule
  \end{tabularx}
\end{table}

\section{Conclusion}

This work demonstrated that real-time digital twins are a viable and effective approach for guiding adaptive scheduling in HPC. We hope this work opens new opportunities for intelligent, simulation-driven scheduling systems that can respond to evolving workload conditions and optimization goals.

\bibliographystyle{plain}  
\bibliography{references}  

\end{document}